\documentclass[11pt,
prd,aps,amssymb,amsmath  ,tightenlines
]{revtex4}
\usepackage{graphics}
\usepackage[pdftex]{graphicx}

\usepackage{amsfonts}
\usepackage{amssymb}

\newcommand{\ket}{\rangle}
\newcommand{\bra}{\langle}

\begin{document}

\title{ Gambling the World Away: Myopic Investors}
 \author{Bernhard K.  Meister}

 \email{bernhard.k.meister@gmail.com}

\date{\today }

\begin{abstract}
\noindent
 Myopic 
  investors  are locally rational  decision-makers but   globally irrational. Their sub-optimal portfolios  lag the market.  
 As a consequence, other market participants are provided with profit opportunities. 
Not  subterfuge but 
constrained optimisation  leads to disparities.  
 Four overlapping examples 
 are given. The first case centres on the difference between  local and global optimisers and their respective Kelly fractions, the second on isolated versus combined optimisation, the third  on the distinction between qualitative and quantitative investor, the fourth on the non-commutative nature of information and the resulting asymmetries. 


\end{abstract}
\maketitle


\section{Introduction}
\label{sec:1a}

\noindent
Myopic behaviour can lead to market distortions.  First an example from the world of lotteries found in Moffitt \& Ziemba\cite{ziemba2018} to develop an intuition. Individuals buy lottery tickets. Innocuously,   they often choose  numbers also preferred by others. As a result, the probability of losing stays unchanged, but when  in luck  winnings have to be shared with a multitude. Individuals picking common numbers underperform, and  numbers chosen in a contrarian fashion add an edge.
Under special circumstances described in the above cited paper 
repeated  lottery drawings turn for some into a `money pump'. This type of arbitrage is not unlimited but real, and can be extended to the financial market as a  whole, since as W. D. `Blackie' Sherrod  said, `if you bet on a horse, that’s gambling.  If you bet you can make three spades, that’s entertainment. If you bet cotton will go up three points, that’s business. See the difference?'
Transaction cost and other frictions can be included without changing the tenor of the argument. It might come as a surprise but in this set up there is no cheating, underhanded dealing, or other type of subterfuge. All is transparent, but still  the clients' (from `cliens', the Latin for `follower' or `retainer')   
portfolios end up underwater. 

Myopia is prevalent in financial markets. Examples are manifold and  include to constrain     asset allocation, to look at assets   in isolation,  to consider restricted information sources or  horizons, and to ignore non-commutativity of investment actions. One of these incongruences will described in each of the next four sections. A section about the paradox of value   followed by a conclusion  rounds of the paper.

\section{
Local versus Global Optimisation}
\noindent
In this section it will be shown that local optimisers invest at a lower Kelly fraction than  diversified global investors. This explains maybe the rise of hedge funds over the last generation. Arguably, it was not their mythical `alphas', which often had  fleeting half-lives, but their ability to allocate capital more widely and therefore more wisely. 

As an example, consider a gambler, who is  only able to bet on one risky double-or-nothing game with win probability   $p$. The optimal Kelly fraction is $2p-1$ for $p> 0.5$.
Compare this to a gambler, who in each round can bet on a host of independent double-or-nothing games  each with win probability   $p$. 
The total optimal investment fraction for the $N$ independent games bettor is 
\begin{eqnarray}
f^*= \frac{p^N-q^N}{p^N+q^N}\geq 2p -1,\nonumber
\end{eqnarray}
where each of the $N$ games is allocated an equal fraction   $f^*/N$ and $q:=1-p$.
In the limit of $N\rightarrow \infty$  the whole capital is employed. 
The fully diversified gambler has an expected return as good  as any other  un-leveraged trading strategy but without excessive risk. This might  come as a belated relief to the founder of a prominent crypto exchanges, which recently imploded, who   stated a preference for maximising the arithmetic over the geometric average.

One could extend this idea to an evolutionary finance setting and imagine different portfolio optimisers with each aiming to maximise the growth of their portfolio. Consumption and other constraints are ignored. What differentiates  groups of investors is  the investment mandate and therefore the range of investable assets, the allowed leverage, and their estimate of the future performance of particular assets. Some investors might be more knowledge or have, correctly or incorrectly, more confidence in their guesses of future returns. The optimal Kelly investment fraction is influenced by all these inputs, and a sensitivity analysis can be found in Lv {\it et al.} \cite{lv2009,lv2010}.
Investors whose estimates are more accurate and make the right choices eventually dominate. Next a section comparing reductionistic and holistic optimisation.

\section{Isolated versus  Combined Optimisation} 
\noindent
This section repeats the argument from the introduction about lotteries but applies it to   financial assets.
Isolated optimisers replicate each other and this leads to herd effects, which can dislocate prices and provide opportunities. As an example, overnight returns in many established stock markets  have  for extended periods been  better than intraday returns. This could be due to the fact that retail investors and institutional managers often form opinions overnight and execute them at the opening of the market, when liquidity is thinner and price impact higher. 
 

In segregated or siloed markets inconsistent models can exist in parallel. In the past, swaptions and caplet models commonly used by market makers in the fixed income markets were not calibrated against the full available data and had different dynamics. As long as churn allowed the capture of bid-offer spreads to mask the inconsistencies, this was able to continue - see the book by Rebonato\cite{rebonato} for details.

Market makers due to risk limits are often forced to accept misconceived pricing models or market inputs due to the general preference for capturing the  bid-offer spread instead of accruing   mis-pricing profits linearly with time, which is more closely associated with proprietary trading or hedge funds. In the case of delta-hedged derivatives the accrual would be of the form $\frac{1}{2}\Gamma(\sigma_{r}^2- \sigma_{i}^2)  dt$ with the difference between the implied volatility $\sigma_{i}$ and the realised volatility $\sigma_{r}$ being the source of the mispricing.  The hedge ratios themselves are dynamic, market and model input dependent,  and differ across market participants. 

Similar `atypical' and exploitable behaviour in crypto markets is often driven by leverage-seeking speculators,  as different ways exist to either establish leveraged short or long positions for capital constrained investors using options. The best strike and maturity to maximise leverage is scenario - stopping time - dependent.  
The leverage based pricing has an effect on implied volatilities and is in contrast to pricing based on expected realised volatilities. The difference of these two pricing models is constantly shifting and exploitable.
The same is true for the futures  market  and should explain part of the observable contango effect in an upward trending market. 
These ideas also play a role in other areas of the crypto market.
 Relative value strategies exist between lending products on over-the-counter as well as DeFi platforms and FX volatilities, since the relative volatilities of interest rates are directly related to the volatilities of FX rates in the spot market.
Another example is scaling laws of market capitalisation, turnover \& volatility of tokens, which are intimately linked and show stability over time. For these scaling laws to retain their form, while there is elevated churn at the lower end with new tokens constantly bubbling up,  the possible dynamics are constrained. Seemingly segregated assets are entwined in an exploitable way. 
Next a section about the disadvantage of qualitative investing.
\section{Qualitative versus Quantitative investors} 
\noindent
Qualitative investors can develop a `passion' for a stock or whole sector 
without being able to quantify  a fair  value.  They follow a `buy  and hold' strategy as long as the enthusiasm persists. The failure    to determine an exit point can result in price distortions, which can easily be viewed as confirmation for the original optimistic view and can lead to irrational exuberance.   
Momentum trading and eventual a mean reversion of the price are consequences. This can be modelled by an Ornstein-Uhlenbeck (OU) process for the underlying market price of risk $\lambda_t$. The drift of the observed asset price $S_t$  oscillates around  the mean reversion value $\hat{\lambda}$ with  the  mean reversion strength of $\kappa$. This is in contrast to a stochastic volatility, a la Heston, with volatility being an OU process.  In this stochastic drift model  the asset price and the market price of risk are given by the following two dynamical equations,
\begin{eqnarray}
\frac{dS_t}{S_t}&=& (r +\sigma \lambda_t) dt+\sigma d W_t,\nonumber\\
d\lambda_t &=&\kappa(\hat{\lambda}-\lambda_t) dt+\hat{\sigma}d\hat{W}_t,\nonumber
\end{eqnarray}
with $W_t $ and $\hat{W}_t$ two independent Brownian motions, and $r$ the short rate and 
$\sigma$ the volatility. This model exhibits momentum as well as mean reversion features as the drift of $\log(S_t)$ oscillates around $ 
 r + \sigma  \hat{\lambda} $.

Another even simpler possibility is to consider the  investment value to be a trending Ornstein Uhlenbeck process,
\begin{eqnarray}
dS_t&=& \Big(\mu -\kappa(S_t-\mu t)\Big) dt+\sigma d W_t,\nonumber
\end{eqnarray}
which again exhibits momentum as well as mean reversion features as the process oscillates around
the trend line given by  $\mu t $.
 This allows the construction of a profitable   strategy that exploits the changing optimal Kelly ratio\cite{lv2010}. 
A section on non-commutative finance follows.

\section{Non-commutative finance}
\noindent
Many phenomena in the natural world are non-commutative, e.g.  rotations, chemical processes    
$\&$ cognition. The
 same is true for finance, since `good earnings by Apple followed by bad earnings by Google' is not necessarily equivalent for the wider market to `bad earnings by Google followed by good earnings by Apple'. The impact of information on prices is non-commutative. Even something as restrictive as order flow is not commutative. A `buy order followed by a sell order' can have a different market impact than the inverse.
Investors understanding the non-commutative nature of markets  have an edge.
 In an idealised setting one can show that for low liquidity and high impact assets this  lead to   arbitrage, which is comparable to a Carnot engine exploiting the temperature differential of two heat baths. Related ideas were discussed in \cite{meister2022}.
 In the next section the paradox of undefined value.
\section{The Paradox of Undefined Value} 
 \noindent
 Instead of `because it is a fraud it has zero value', the statement should be `because it is a fraud it has an arbitrary value'. Something analogous is true in mathematics, where an infinite sum can have a well defined value, or in the case of the divergent Grandi's series, i.e. $\sum_{k=1}^{\infty}(-1)^k=1-1+1-1,...$, be ill-defined.   As a consequence, the value of an ill-defined series of pay-outs is inherently flexible. 
  A conclusion rounds of the paper.
\section{Conclusion: Why Arbitrage has not been Arbitraged away?}
\noindent 
Much has been written about arbitrage.  To paraphrase Mark Twain, `the reports of its death are greatly exaggerated'. 
What is arbitrage and why does it persist?
We do not consider here the 
 financial mathematics definition of  `free lunch with vanishing
risk', but instead a looser  definition based on excess return without excessive risk.
The  clash of different views of the world is expressed through prices. A struggle often ensues and one view of the world eventually prevails. The gold to silver ratio differed between China and much of the rest of the world during the 16th to 17th century. This led to a brisk trade, a build up of tension and an eventual exchange rate adjustment. Exploiting the discrepancy took time and was not without operational risk in the form of shipwrecks, pirates and other imponderables.  Capital, which has a cost and constraints, was required to exploit the opportunity. 
Arbitrage can be compared to diving. If a diver is forced to stay under water too long the air runs out, i.e. capital and access to leverage disappear.
 In its various forms, arbitrageurs profit  from the existence of inconsistent prices, and the gradual or instantaneous shift towards consistency, even if  achieving consistency in one part of the market can open up inconsistencies elsewhere.  
How is an arbitrage opportunity exploited? A ‘money pump’ can be constructed from the idiosyncrasies of market participants, which are often locally rational but due to myopic tendencies explained above fail to avoid pitfalls. For other examples of `money pumps' see \cite{meister2016,meister2022}.   
As a consequence, arbitrageurs normally outperform, while    retail and many institutional investors  consistently  underperform, e.g.  `where are the customers' yachts' was a  self-deprecating book title.  The reason has to be a little bit subtler than market frictions, even if this adds to the malaise. The hypothesis of the paper is instead that there are  exploitable cognitive   or other failures. Relative performance is in many cases as important as absolute performance, since as J.M. Keynes wrote,  `worldly wisdom teaches that it is better for reputation to fail conventionally than to succeed unconventionally'. 

\hspace{-.538cm}The author wishes to express his gratitude to D.C. Brody and H.C.W. Price for  stimulating discussions.




%

 

\vspace{-.030288cm}
\end{document}